\documentclass{article}
\pdfoutput=1
\usepackage{aaai17}
\nocopyright
\usepackage{url}
\usepackage{amsfonts}
\usepackage{amssymb}
\usepackage{amsmath}
\usepackage{graphicx}
\usepackage{longtable}
\usepackage[symbol]{footmisc}

\title{The Device War \\ The War Between IOT Brands In A Household}

\author{Marius C. Silaghi\footnotemark, Arianit Maraj$^*$, Timothy Atkinson$^*$\\
Florida Intitute of Technology\\
msilaghi@fit.edu,amaraj@my.fit.edu,atkinsot1999@my.fit.edu
}
\date{}

\begin{document}
\pdfinfo{
	/Title (Learning strategies for power attacks on Direct Wi-Fi 802.11 group formation)
	/Author (Marius C. Silaghi, Arianit Maraj, Timothy Atkinson)
	/Keywords (IoT, Bayesian Neyworks, Wi-Fi Direct, Wi-Fi P2P, ad-hoc)
}
\maketitle
\footnotetext[1]{Contact Authors: Marius C. Silaghi $\langle$msilaghi@fit.edu$\rangle$, Arianit Maraj $\langle$amaraj@my.fit.edu$\rangle$, Timothy Atkinson $\langle$atkinsot1999@my.fit.edu$\rangle$}

\begin{abstract}
Users buy compatible IOT devices from different brands with an expectation
that their cooperation is smooth, but while function may superficially look friendly,  cohabitation can subversively cause early battery depletion in competitor devices.
	
	The Wi-Fi Direct standard was introduced with the intention of simplifying peer-to-peer connections in home applications while helping devices to save power through centralization of effort into a single group owner device negotiated on start-up.
	Attacks on the group formation stage can be based on manipulating a victim device to frequently end up being assigned the group owner function, thereby depleting its batteries at faster rates than its peer devices. This manipulation is made easy by the group formation mechanism adopted by the standard.
	
	We show that group formation procedures could be better secured with features ensuring fairness by relying on commitments and by learning about the behavior observed for peer devices in the past.
	Simulations are used to quantify the resistance achieved against several attack strategies.
\end{abstract}

\section{Introduction}

Mechanisms to mitigate attacks on devices in smart homes and other IoT applications are introduced based on learning of peer behavior from sequences of outcomes from Wi-Fi Direct group formation processes, and by enforcing fairness with cryptographic commitments.
The recent Wi-Fi Direct standard was developed as an alternative to the previous Wi-Fi modes, of which the most well known are the infrastructure mode and the adhoc mode. While the infrastructure mode only allows devices to communicate in the presence of an Access Point~(AP), the adhoc Wi-Fi mode 
causes intensive power consumption by devices
as they have to continuously
advertise themselves. 
In comparison, the new Wi-Fi Direct mode comes with 
standard set-up procedures, after which only one device needs to consume power for broadcasting maintenance signals. This device is called group owner~(GO), and it is appointed via a negotiation protocol.

\paragraph{Motivation}
Devices in a smart home may come from different vendors and are
allowed to communicate for achieving higher level tasks, such as regulating light and sound
levels in a larger space. Devices from some vendors may exploit devices of a competitor by coaxing them into taking a GO function in a disproportionate way with the intention to deplete their batteries faster.
This in turn will cause a negative impact on the user perception about those competitors.
More dangerously, a device that has been taken over by an attacker can be used as a proxy in an
attack on the batteries of a more sensitive device, such as a door-keeper, by similarly manipulating it into becoming GO.\\

\noindent
{\bf Definition}
{\it	A {False Friends Battery Depletion (FFBD)} attack is when a victim device is intentionally induced into depleting its battery by volunteering services.
}

E.g., this could occur while attackers disproportionately avoid such duties, or do not need those services.

\paragraph{Technology}
The standardized negotiation protocol does not account for security attacks.
Each device configures a level of preference for becoming GO, as an integer Intent Value (IV) between 0 and 15. A preference of 15 communicates a requirement of being GO, without which the negotiation fails. In case both preferences are between 0 and 14, the device with the highest value becomes GO. For the case where preferences are equal, the contacting device includes a Tie-Breaker Bit (TBB) that is supposed to be flipped in subsequent requests. The bit is flipped in the reply, and the device sending the TBB 1 will become GO. Finally,
devices can abandon the group formation for any reason.

The standard itself does not specify how devices should select their GO preferences, and there are no mechanisms to verify or ensure that TBBs
flip between sessions, as recommended. A device that wants to avoid being GO, can set IV to 0, and also ground the TBB.

\paragraph{Approach}
To address
False Friend Battery Depletion attacks we propose 
mechanisms based on 
learning peer profiles from their behavior over time
and using
secure bit commitments for flipping TBBs, to improve the detection of manipulating devices.

In the next section we introduce background about Wi-Fi Direct and its group formation sub-protocol,
as well as on the various competing techniques for secure bit commitment 
and coin flipping.
After describing related work that focuses on other attacks on Wi-Fi Direct and group formation, we will delve into strategies for learning and adjusting to attacking peers. In the section Exploiting Commitments we describe the proposed integration of fair bit flipping techniques into the group formation protocol.
We end by describing experimental results based on simulations, confirming the robustness brought by the proposed techniques with respect to various attack scenarios.

\section{Background}

This section introduces the recent Wi-Fi Direct technology, focusing on its
group formation subprotocol, and research on intelligent strategies for power consumption reduction, as well as power consumption attacks.

\subsection{Wi-Fi Direct}
The 802.11 standard allows vendors to extend the specification through their Information Element (IE) Frame, as we also propose here. The IE has a general format consisting of one-octet element ID field, one-byte length field, three-bytes OUI field, one-byte OUI type and up to 251 bytes of payload~\cite{IEEE_standard}. 
Wi-Fi Direct defines the P2P IE Frame using attributes which are composed of a one-byte attribute ID, a two-bytes attribute length, and attribute data.  Multiple attributes can be placed within a single IE Frame, and compliant vendors are required to ignore attribute IDs they do not understand (including attribute IDs marked as reserved in the current standard).  New attributes can be used for extensions as proposed here.

This protocol allows communication among peers within a single group~\cite{Alliance}.  The roles of GOs and clients are defined during the group formation process. The P2P GO implements an AP-like functionality~\cite{Altaweel2017EvilDirectAN}, with maintenance and advertisement performed through periodically sent beacon frames, increasing its power consumption.  After the election process, the role of GO or client remains unchanged during the entire group session. When the GO owner, for any reason, leaves the group, the devices become disconnected.

\subsection{Group formation process}

There are three methods for creating a group: standard, persistent, and autonomous. However, our focus in this paper is only on the standard group formation method. 

\subsubsection{Standard Group formation}

\begin{figure}[!t]
	\includegraphics[width=0.8\columnwidth]{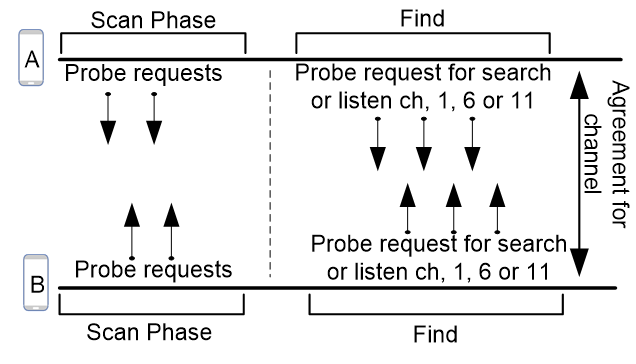}
	\centering
	\caption{P2P Device discovery process}
	\label{fig:Figure2}
\end{figure}

The device discovery phase consists of two sub-phases: scan and find. During the scan phase, devices try to find other devices/groups or Wi-Fi networks, also locating the best channels for establishing a group. Devices cannot reply to request frames when they are involved in scanning. During the find phase, the device selects one of the channels 1, 6, or 11 in the 2.4GHz or 5GHz bands, as Listening Channel. After this, the device alternates between a scanning phase by sending Probe Requests in each of these channels; and a listen state, in which it listens on a channel for Probe Requests frame in order to respond with Probe Response frames. The discovery phase algorithm is illustrated in Figure~\ref{fig:Figure2}. 

\begin{figure}[!t]
	\includegraphics[width=0.99\columnwidth]{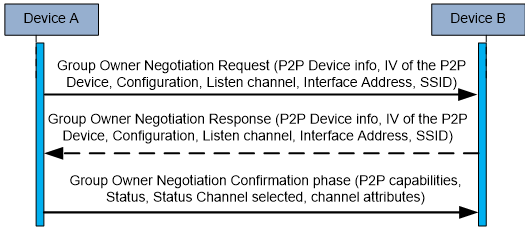}
	\centering
	\caption{Group owner negotiation process}
	\label{fig:Figura3}
\end{figure}

Once a device has found another device that it wants to contact, and which is not yet involved in a group, it starts the so called Group Owner Negotiation phase.
The details regarding this three way handshake communication are illustrated in Figure~\ref{fig:Figura3}. 
When the target device is already in a group, the owner of that group is contacted, skipping the GO negotiation phase.

\subsection{Power consumption}

Since the GO device might be battery powered, its energy consumption is critical~\cite{6876837,liao2017designing}.
To help reduce consumption, the protocol can define an availability period of the GO, called  Client Traffic Window.
Power management for GO consists of delivery mechanisms defined for Power Save and Wi-Fi multimedia power save (WMM-PS), but there also exist power saving mechanisms that allow for a GO to be sometimes absent: the Opportunistic Power Save (OPS), and the Notice of Absence (NoA).

\subsection{Secure Commitments}
The breaking of ties between peers with identical {\em intent value} to become GO
is performed in the P2P Direct standard by trusting the contacting peer
to locally generate a random bit. However, this trust is not rational in the light
of FFBD attacks. 

One of the solutions we propose is to
use
secure coin flipping over the network. Many of these techniques employ bit commitment protocols.
An example of such a technique is the simple bit commitment protocol in the random oracle model~\cite{camenisch2018wonderful}. Assume an agent $A$ wants to commit to a bit with value $b$. $A$ generates a sufficiently large random number $R$,
and then publishes the value $c=H(R,b)$ where $H$ is a secure message digest function such as SHA-2 or Whirlpool~\cite{rfc}.  The commitment is {\em binding}, in the sense that it is difficult for $A$ to find other values for $R$ and $b$ that generate the same commitment $c$. If $R$ is large enough, the commitments is also {\em hiding} in the sense that an attacker would not be able to efficiently try sufficiently many values of $R$ and $b$ in order to find one that matches $c$.
Bit commitment techniques have also been proposed based of pseudo-random number generators as well as on other security concepts~\cite{naorprng}.

Coin flipping over the network~\cite{blum1983coin} by two agents $A$ and $B$ can be performed by
having $A$ commit to a bit $b_A$ before $B$ publicly discloses a bit $b_B$.
When $A$ receives $b_B$ then it also reveals $b_A$. The result of the coin flip is the exclusive OR of the bits, $b_A \otimes b_B$.

\section{Related Work}

Due to the open access, reduced security infrastructure, and broadcast nature of communication, Wi-Fi direct is threatened by different cyber-attacks~\cite{shen2016secure}. 

Two commonly addressed attacks are Impersonation and Denial of Service (DoS).\\ 
{\it Impersonation.} The attacker can impersonate a MAC address and then change the Wi-Fi channel. When a device attempts to join a legitimate network, it can be deceived into connecting to a device of an attacker.\\
{\it DoS.} The aim of the DoS attack is to overload or eventually completely crash a system or network, by flooding it with useless traffic. Many smart home devices are battery operated, thus flooding these devices with requests can lead to an energy depletion attack, preventing legitimate users from having access to the system~\cite{7977239}. The battery exhaustion attack is a classic DoS attack which tries to reduce the battery life of mobile devices~\cite{nash2005towards}, in particular to reduce the battery of the GO. Battery exhaustion attacks have been studied in Wireless Sensors Networks and Internet of Things~\cite{4431860}, as well as on battery-powered mobile devices~\cite{1276868}.

\section{Learning Peer Behavior}

\newcommand{\SA}{\ensuremath{A_{SA}}}
\newcommand{\MA}{\ensuremath{A_{MA}}}
\newcommand{\F}{\ensuremath{A_{F}}}
\newcommand{\BA}{\ensuremath{A_{BA}}}
\newcommand{\AL}{\ensuremath{A_{AL}}}

\newcommand{\VH}{\ensuremath{V_{H}}}
\newcommand{\VB}{\ensuremath{V_{B}}}
\newcommand{\VA}{\ensuremath{V_{A}}}
\newcommand{\VS}{\ensuremath{V_{S}}}
\newcommand{\VL}{\ensuremath{V_{L}}}

Detecting and handling graciously a FFBD attack requires complex reasoning or a robust classifier.
Designing a set of logic rules leading to a rigid classification of a peer as attacker is possible but fragile due to uncertainty.
A probabilistic classifier implemented as a Bayesian Network or an Artificial Neural Network can more robustly be used to associate a FFBD attack to a posterior probability, given a history of communication.

Once an attack probability is associated with each peer profile, this can be communicated to the device owner (or manufacturer). Further, automatic acceptance or rejection of a connection as GO with a peer can be made using a utilitarian approach, based on the current batteries levels, a fairness measure, and a parameter of risk adversity that can be configured.

We propose to consider that a GO device has {\bf prematurely quit} a group if it either quits before the end of a group formation negotiation where it would end up as GO, or it quits before the group can be used to perform any concrete application-level task.

Let us now describe the investigated solution based on a Bayesian Network. The module is assumed to maintain a profile for each peer with which it creates groups. The profile stores statistics used as features of the learning process. These statistics can be maintained over a sliding time window (a month, in our experiments). 
To maintain the sliding window,
the statistics for a time window can be assembled dynamically from smaller buckets (we use one bucket per day), such that for each new bucket added, the oldest bucket is removed from the totals. Identified features, computed by a device D for each peer over the sliding window, are:
\begin{itemize}
	\item {\bf iGO}:
	percentage of GO negotiations with peer where D was elected GO. The possible values, as used in our experiments, are: low (\VL) for 0-25\%, small (\VS) for 25-40\%, average (\VA) for 40-60\%, above (\VB) for 60-75\%, and high (\VH) for 75-100\%.
	\item {\bf pGO}: 
percentage of GO negotiations where peer is GO and quits prematurely,
discretized as for iGO.
	\item {\bf tGO}:
	The percentage of communication time with peer in which D was GO.
	\item {\bf Data}:
	number of GO negotiations with the peer. The domain of Data is discretized as "insufficient knowledge" (I) below 10 rounds, "some knowledge" (S) below 100 rounds, and "reasonably informed"~(R).
\end{itemize}

\begin{figure}[t!]
	\centering
	\includegraphics[width=0.4\columnwidth]{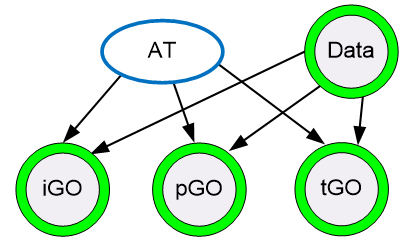}
	\caption{Bayesian Network. 
	 }
	\label{fig:BN}
\end{figure}

\noindent
The above features are used by the device learning module to output peer characterization values:

\begin{itemize}
	\item 
	False Friend Battery Depletion attack ignorance: the current ignorance of the device as to whether there exists sufficient data to judge whether the peer is an FFBD attacker, in the sense of the Dempster-Shafer theory~\cite{shafer1976}. 
	It is mapped from the values of the {\it Data} feature.
	\item 
	Peer Fairness ($PF$): how fair has been so far the GO time distribution with this peer.
	It is estimated as the percentage of communication time with peer in which D was GO, namely $tGO$. E.g., ratios below .6 are considered fair.
	\item 
False Friend Battery Depletion attack (AT) probability: the  probability associated with being under an FFBD attack from peer,
estimated using the Bayesian Network in Figure~\ref{fig:BN},
where the thick margin circles are evidence variables, and the conditional probability tables (CPT) are shown in Figure~\ref{fig:CPTs}.
\end{itemize}

\noindent	
FFBD attacks can be at different intensity.
The attacker models represented by the multi-value random variable AT, are classify as (see  Figure~\ref{fig:CPTs}): 
\begin{itemize}
\item "Strong attacker" (\SA), where the percentage of negotiations where peer attacks is high~(\VH)
\item "Medium attacker" (\MA), where the percentage of negotiations where peer attacks is above average (\VB)
\item "Fair" (\F), where the percentage of negotiations where peer attacks is average (\VA)
\item "Better than average" (\BA), where the percentage of negotiations where peer attacks is small~(\VS)
\item "Altruist" (\AL), where the percentage of negotiations where peer attacks is very low (\VL)
\end{itemize}
	
\begin{figure}[!t]
	\tiny
\begin{tabular}{|ll|}\hline
Data=R&	\begin{tabular}{|l|l|l|l|l|l|}\hline
		AT & P(\VL) & P(\VS) & P(\VA) &P(\VB) &P(\VH)
		\\\hline
		\SA & 0.05 & 0.1 & 0.2 & 0.2 & 0.45 \\\hline
		\MA & 0.05 & 0.1 & 0.2 & 0.45 & 0.2 \\\hline
		\F  & 0.1 & 0.2 & 0.4 & 0.2 & 0.1 \\\hline
		\BA & 0.2 & 0.4 & 0.2 & 0.1 & 0.1 \\\hline
		\AL & 0.4 & 0.2 & 0.2 & 0.1 & 0.1 \\\hline
	\end{tabular}\\\hline
Data=S &	
\begin{tabular}{|l|l|l|l|l|l|}\hline
	AT & P(\VL) & P(\VS) & P(\VA) &P(\VB) &P(\VH)
	\\\hline
\SA & 0.14 & 0.14 & 0.14 & 0.22 & 0.36 \\\hline
\MA & 0.13 & 0.13 & 0.20 & 0.34 & 0.20 \\\hline
\F & 0.13 & 0.20 & 0.34 & 0.20 & 0.13 \\\hline
\BA & 0.20 & 0.34 & 0.20 & 0.13 & 0.13 \\\hline
\AL & 0.36 & 0.22 & 0.14 & 0.14 & 0.14 \\\hline
\end{tabular}\\\hline
Data=I&	
\begin{tabular}{|l|l|l|l|l|l|}\hline
	AT & P(\VL) & P(\VS) & P(\VA) &P(\VB) &P(\VH)
	\\\hline
	\SA & 0.17 & 0.17 & 0.17 & 0.24 & 0.25 \\\hline
	\MA & 0.15 & 0.15 & 0.23 & 0.24 & 0.23 \\\hline
	\F & 0.15 & 0.23 & 0.24 & 0.23 & 0.15 \\\hline
	\BA & 0.23 & 0.24 & 0.23 & 0.15 & 0.15 \\\hline
	\AL & 0.25 & 0.24 & 0.17 & 0.17 & 0.17 \\\hline
\end{tabular}\\\hline
\end{tabular}
	\centering
	\caption{Conditional Probability Tables of the leaf nodes, built as a definition of the attacker types.
	}\label{fig:CPTs}
\end{figure}

The prior probability of other devices being attackers can depend on the manufacturer of the other devices.
In our experiments this multivalued random variable prior is the vector of 5 values $[0.15, 0.2, 0.45, 0.1, 0.1]$, and in practice can be adjusted based on inputs from the end-used concerning their happiness with the relative battery performance of the device.
The FFBD attack probability for a peer is checked each time the device becomes GO, and if $PF > 0.6$, computing:

$P(AT|iGO,pGO,tGO,Data)$\\
\strut\hfill$= \alpha P(iGO,pGO,tGO|AT,Data)P(AT)$

\section{Exploiting Commitments}

\label{sec:commitments}
In addressing FFBD attacks, solutions have to consider the trade-off between efficiency and security. The introduction of a fair coin-flipping 
mechanism provides a learning component with strong evidence of manipulation intention on the part of one's peer in a group owner negotiation, when this quits prematurely.
The coin-flipping solution based on the random oracle model is a choice with small computation requirements.

While the $IV$ value of a device may depend on the peer with each it communicates,
the TBB its supposed to be independent and a commitment to it can be advertised in Probe Request frames. A tentative IV value can also be advertised with the probe, but it is understood that the device may change the value based on the peer identity.
In a version that we propose along with our learning strategies,
each peer $A$ of the GO negotiation will commit to its current $TBB_A$ by sending
its commitment $H(R_A|IV_A|TBB_A)$ inside a new type of Information Element broadcast with each Probe Request, called Tie Breaker Bit Commitment (TBBC) IE.

\begin{figure}[!t]
	\includegraphics[width=0.99\columnwidth]{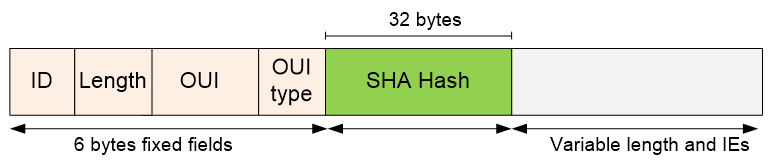}\centering
	\caption{Proposed Vendor IE frame format}\label{fig:Figure6}
\end{figure}

The number of bytes of overhead required for a SHA-256 hash is 38, namely 32 bytes for the hash and 6 bytes for the IE header, if the message is implemented as a {\it Vendor IE} (see Figure~\ref{fig:Figure6}). 
Instead, the overhead is 35 bytes if the message uses one of the {\it Reserved P2P attributes}.
Further, for compatibility, the GO Negotiation Request may contain a random $TBB'_A$ that is not correlated with the $TBB_A$ committed to.
 This adds 4 bytes: 3 as overhead for a P2P attribute header, and 1 byte of payload for the TBB bit.

The device $B$ deciding the identity of the GO will then use as tie breaking value the 
result $TBB$ of the computation $TBB=TBB_A~XOR~~TBB_B$.
Then, device $B$ opens its commitment using the GO negotiation response to which it can add a P2P IE attribute for opening the commitment by stating its $R_B$. It will
also communicate the $IV_B$ actually used in the decision.
Differences between this $IV$, $IV_B$ and past values are also reliable features usable by the learning system to build a profile for device $B$.

\begin{figure}[!t]
	\includegraphics[width=0.90\columnwidth]{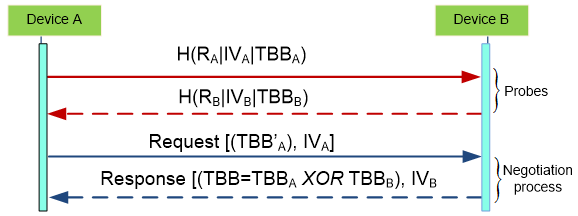}
	\centering
	\caption{Proposed probe and negotiation process}
	\label{fig:Figure7}
\end{figure}

An advantage of this solution is that the need of GO negotiation confirmation may be reduced and eventually removed from the protocol, shortening the setup delay in future versions of P2P Direct, which will improve its usability to applications sensible to start-up latency, such as vehicle-to-vehicle communication.

\paragraph{Commitment In GO Negotiation}

An alternative approach is to integrate the TBB commitment fully in the GO negotiation process.
The device A integrates the commitment $H(R_A|IV_A|TBB_A)$ in the GO negotiation request.
Further, the device $B$ generates both sets of responses for the cases where a GO is assigned to device A and to device B. Both outcomes are presented in the GO negotiation response, together with $TBB_B$ and $IV_B$.
In the end, the device $A$ returns the GO negotiation confirmation with the selected result, and opening the commitments for $TBB_A$ and $IV_A$.

While commitments do not allow cheating with TBBs, attackers can still quit prematurely.

\begin{figure}[!t]
	\includegraphics[width=0.9\columnwidth]{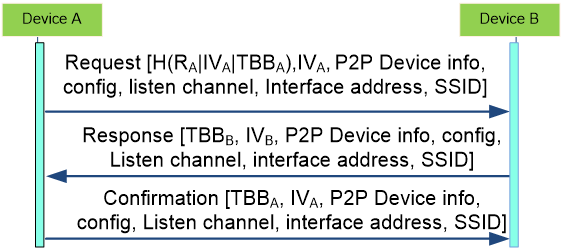}
	\centering
	\caption{Commitment in the GO negotiation}
	\label{fig:Figure8}
\end{figure}

\section{Experiments}

A simulator was implemented and used to run experiments with 2, 5, and 10 devices.
Each device is assumed to start with a battery that can last 365 days in the non-communicating mode. 
The simulation assumes that the non-communicating mode uses 1 energy unit per second. The client mode uses an additional energy unit per second. The GO mode uses 10 additional energy units per second, besides the unit used by the non-communicating mode.
The IV is always set to 0 to focus on the tie-breaking algorithm.
In all the simulations, the device 0 may be a victim of FFBD attacks by a subset of the remaining devices. 
The efficiency of the tested algorithms is computed as the time until the victim battery depletes. As expected, the battery lasts longer when the victim defends itself by either rejecting connections from detected attackers classified using the Bayesian Network (``L''), using secure bit commitment (``C''), or both (L+C), rather than the standard method (``S''). Each point is averaged over 10 runs performed with different random seed values.
Simulations where the attackers also quit prematurely, rejecting assigned GO roles and retrying connections are marked with ``R''.
A victim drops a GO Request from peers detected as attackers, if their peer fairness is $PF>60\%$.

\begin{figure}[!t]
	\includegraphics[width=0.95\columnwidth]{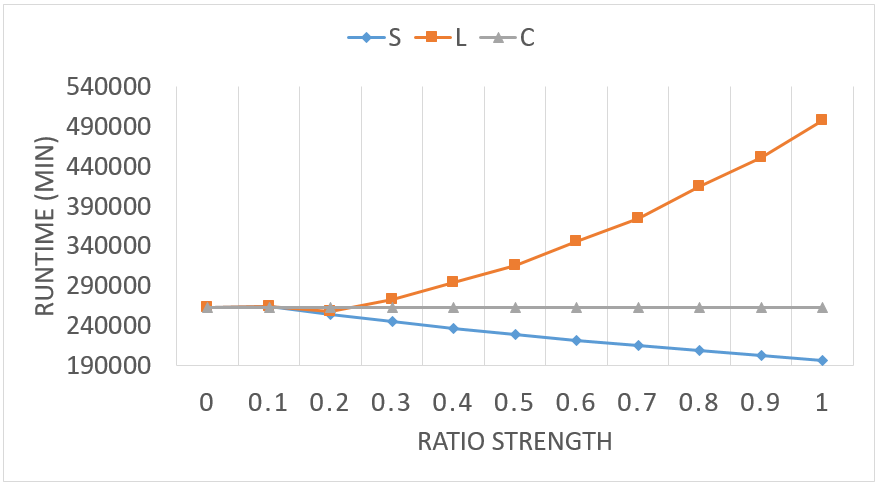}
	\centering
	\caption{Effect of various TBB attack strengths}
	\label{fig:VarStrength}
\end{figure}

Figure~\ref{fig:VarStrength}  describes the effect of the {\em TBB attack strength} (percentage of manipulated TBBs). 
In this set of experiments, there are only 2 devices, namely the victim and the FFBD attacker. 
For each simulation, the attacker follows a repeated communication schedule with groups lasting for 1 minute every 6 minutes.
As noticed, the attacks are detected by the Bayesian Network starting at the strength of 30\%, while commitments can enforce fair behavior.

\begin{figure}[!t]
	\includegraphics[width=0.95\columnwidth]{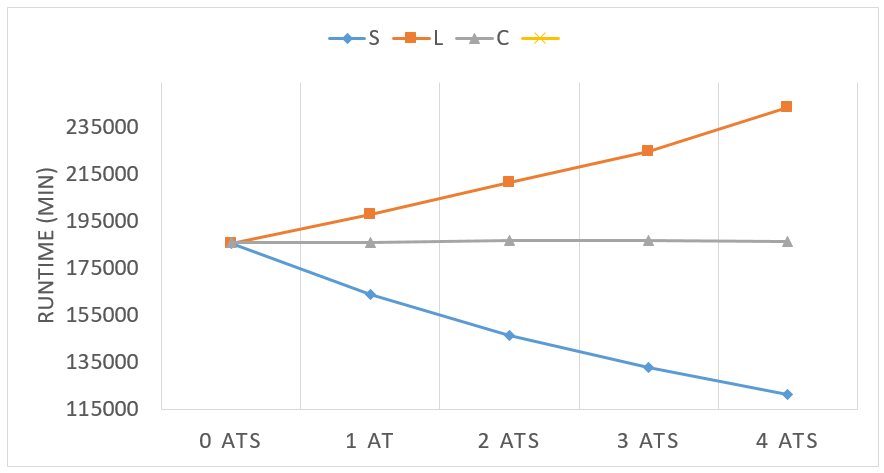}
	\centering
	\caption{Effect of ratio of attackers with 5 devices}
	\label{fig:5Dev}
\end{figure}

\begin{figure}[!t]
	\includegraphics[width=0.95\columnwidth]{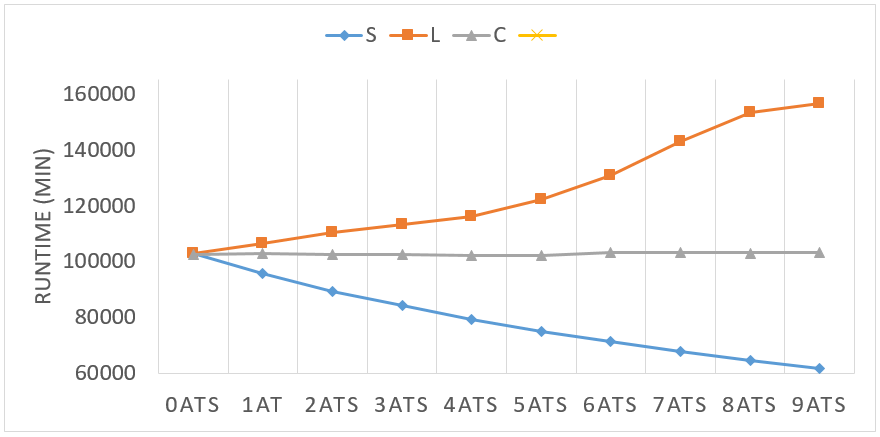}
	\centering
	\caption{Effect of ratio of attackers with 10 devices}
	\label{fig:10Dev}
\end{figure}

The Figures~\ref{fig:5Dev} and~\ref{fig:10Dev} describe the performance with 5 and 10 devices, respectively. 
The devices follow a repeated communication schedule with groups lasting for 1 hour every 12 hours.
For these sets of experiments, a fraction of the devices act as attackers by manipulating the TBB in each group formation process that they initiate. It is also observed that the battery of the victim will last longer with method L and many attackers since the victim device will detect and avoid unfairly communicating with attackers in GO mode, saving more battery.

\begin{figure}[!t]
	\includegraphics[width=0.95\columnwidth]{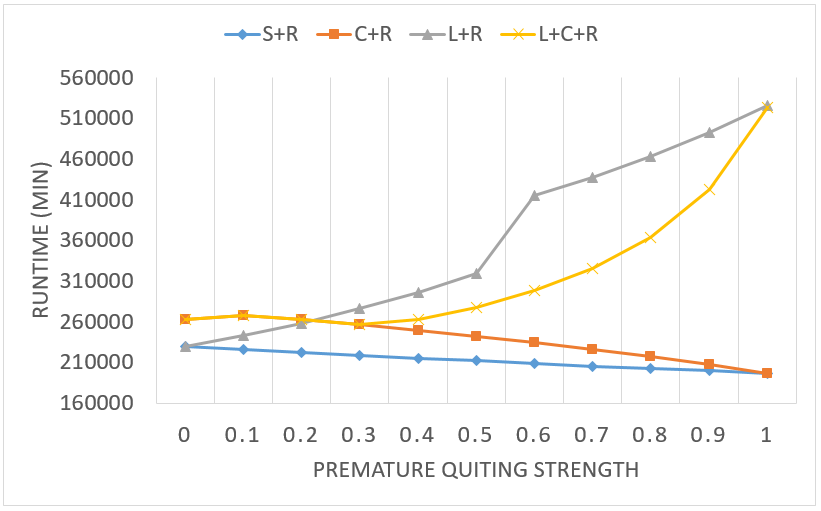}
	\centering
	\caption{Effect of various R attack strengths}
	\label{fig:2R}
\end{figure}

Figure~\ref{fig:2R} describes the effect of the {\em R attack strength} (percentage of negotiations where the attacker quits prematurely when becoming GO). All points are computed at a TBB attack strength of 50\%, while the R attack strength varies. While it is observed that, as expected, secure bit commitment cannot defend from a strong R attack, learning allows for longer battery life of victims as they drop communication with attackers.

\paragraph{Future Work}
Developing a utility driven behavior for devices is the next logical
extension of this work. The utility that a device is expected to associate to a state of the world is basically the utility it contributes to the {\em brand} of the manufacturer. This utility reflects the way in which the profits of the manufacturer are impacted by the performance of its devices, through their effect on the image of the brand:
\begin{itemize}
	\item
	Customer happiness with respect to battery life.
	\item
	Customer happiness related to device inter-operability and availability.
\end{itemize}
These two components of utility provide incentives towards contradictory behaviors, and the optimal trade-off depends on their actual values, relation that can be obtained as input from the end-customer.

To avoid letting an attacker deplete a victim's battery by keeping it as GO after connecting to it while it was already GO as a result of a negotiation with a third party, as future work we will also investigate deadlines after which GOs decides to close a group. The deadlines can be set with an alarm started at the GO negotiation time, or when the first client quits the group.

\section{Conclusions}

The False Friend Battery Depletion (FFBB) attack is identified as a potential hazard against devices
in smart homes, where battery-powered victims are coaxed into volunteering to support attacker devices, as intensive power consuming group owners, reducing their own availability to end-users and damaging their brand reputation.
We address FFBD attacks based on manipulating TBBs and on premature quitting.
A set of solutions is proposed for addressing FFBD attacks, using secure bit commitments, learning statistical profiles of peers, and classifying them using Bayesian Networks.
The solutions based on secure bit commitments prove to be very robust, but would require changes in the Wi-Fi Direct standard group formation protocols, changes which are unlikely to occur soon given that many devices already support the current standard.
The results based solely on learning statistical profiles and detection using Bayesian Networks were shown to also work very well, being a valid alternative where attacks are
detected and automatically denied.

\bibliographystyle{aaai}
\bibliography{wifidirect}

\end{document}